# Real-Time Monitoring of Fluorescence *in situ* Hybridization (FISH) Kinetics

Nadya Ostromohov [1,2,§], Deborah Huber [1,§], Moran Bercovici [2] and Govind V. Kaigala [1]*

[1] IBM Research—Zurich, Säumerstrasse 4, 8803 Rüschlikon, Switzerland
[2] Faculty of Mechanical Engineering, Technion – Israel Institute of Technology, Haifa 3200003, Israel

## Abstract

We present a novel method for real-time monitoring and kinetic analysis of fluorescence *in situ* hybridization (FISH). We implement the method using a vertical microfluidic probe containing a microstructure designed for rapid switching between a probe solution and a non-fluorescent imaging buffer. The FISH signal is monitored in real time during the imaging buffer wash, during which signal associated with unbound probes is removed. We provide a theoretical description of the method as well as a demonstration of its applicability using a model system of centromeric probes (Cen17). We demonstrate the applicability of the method for the characterization of FISH kinetics under conditions of varying probe concentration, destabilizing agent (formamide) content, volume exclusion agent (dextran sulfate) content, and ionic strength. We show that our method can be used to investigate the effect of each of these variables and provide insight into processes affecting *in situ* hybridization, facilitating the design of new assays.

## Introduction

Fluorescence *in situ* hybridization (FISH) refers to a group of methods widely used for the detection, visualization and analysis of specific nucleic acid sequences in cells, tissue sections, and whole organisms.[1,2] FISH is the gold standard cytogenetic technique commonly used for the detection of chromosomal and transcriptional abnormalities for research and medical diagnostics purposes. Its ability to provide a detailed spatial analysis of gene expression and chromosomes at the single cell level, to identify the cell type, and to detect presence of viral genomes in the analyzed sample has led to its extensive use in a wide range of applications.[1,3–7]

FISH reactions are typically performed using a bench-top assay by pipetting the FISH hybridization mix onto the cytological sample, in which the transport of probes into the cell is primarily diffusion-based and hence is slow.[1,8] This leads to typical incubation times of up to 16 hours for low copy number targets[3,9] and 48 to 94 hours for entire genome hybridization.[4]

Once hybridization is complete, FISH signals are analyzed using an end-point analysis after a stringent wash of the hybridization buffer containing the fluorescent probes from the surface.[1] An accurate quantification of the hybridization rates in FISH reactions requires the monitoring of the process in real time. However, such monitoring is challenging owing to the background signal of a high concentration of probes present on the surface at all times throughout the experiment. Furthermore, standard techniques for the evaluation of molecular interactions and reaction kinetics, such as surface plasmon resonance (SPR),[10] or isothermal titration calorimetry (ITC),[11] are not compatible with intracellular reactions as they require either direct immobilization of the reactants on the surface, reactions occurring at distances smaller than 200 nm from the surface,[12] or a uniform solution environment.[13]

While hybridization rates obtained in solution or using surface-based methods may provide general guidelines or relative results for a given set of complementary sequences, they cannot assess hybridization rates for large molecules such as chromosomes or an entire genome.[4,14] Further, these methods do not account for factors such as intracellular and intranuclear molecular crowding,[15] spatial arrangement and target accessibility,[16,17] and the significantly slower diffusional transport due to a dense nuclear and chromatin structure,[18,19] and thus cannot provide the appropriate reaction rates of the *in situ* reaction.

In recent years, there is a growing interest in a more precise evaluation of *in situ* hybridization rates to achieve probe designs with improved sensitivity and specificity and shorten the typically long assay times.[20] Multiple computational models were proposed to predict the hybridization kinetics of FISH probes, including accounting for nucleic acid hybridization thermodynamics, rRNA and genomic DNA accessibility,[17,21,22] and the effects of destabilization agents (e.g. formamide) on probe dissociation.[23] However, models able to accurately predict the optimal hybridization conditions for a given probe do not yet exist.[20] Several experimental studies aimed at improving the hybridization rates as well as at optimizing existing assays were also performed recently. Matthiesen *et al.* designed a non-toxic ethylene carbonate (EC)-based hybridization buffer, accelerating the FISH reaction and resulting in notable signals after 60 min of hybridization,[24] Stoecker *et al.*[25] used double-labeled probes to enhance the signal and shorten the required hybridization time, and Shaffer *et al.*[26] optimized the fixation of cells and increased the probe concentration to achieve 10-fold shorter hybridization times of RNA FISH. Other methods suggested for measuring intracellular kinetics include the use of optically switchable probes, such as molecular beacons[27] and temperature oscillation optical lock-in (TOOL) microscopy requiring Förster resonance energy transfer (FRET) probes.[15] However, despite their fundamental importance to FISH assay development, there are currently no techniques for the quantification of the hybridization rates that can be readily applied to any probe or target. Thus the design procedures of FISH probes and assays remain mainly empirical, and the hybridization conditions for a given probe are estimated experimentally using an end-point analysis of the signal.[20,21] This remains a highly complex, labor-intensive and time-consuming task when multiple effects and conditions are to be tested.

Here we present a new method for designing FISH assays that are capable of accurately characterizing the hybridization kinetics for any given set of probes or target sequences. Our method is based on a vertical microfluidic probe (MFP), able to deliver reagents of interest to localized regions on a surface.[28] We demonstrate that by rapidly alternating between delivery of the probe solution and of a non-fluorescent imaging buffer, we are able to monitor the FISH signal in real time. The convection-based transport using the MFP enables delivery of a constant concentration of probes to the cells and rapid removal of any signal associated with unbound probes while the signal is being recorded.[29,30] This allows real-time monitoring and kinetic analysis of the signal inside the cell. Using centromeric probes (Cen17) as a model system, we demonstrate the applicability of our method for measuring variations in the hybridization rates in response to changes in probe concentration, ionic strength, formamide concentration, and volume exclusion agents content. To the best of our knowledge, this is the first method of its kind, capable of direct real-time monitoring of *in situ* hybridization kinetics without the need for specialized nucleic acid probes. We believe our method will be useful for the design of new as well as the optimization of existing FISH assays by providing new insights on the *in situ* hybridization (ISH) process.

# Experimental

## Experimental Setup

We performed all experiments using an inverted microscope (Eclipse-Ti, Nikon Instruments, Melville, NY), equipped with an Orca Flash 4.0 camera (Hamamatsu Photonics, Hamamatsu, Japan), an LED white light source (Sola, Lumencor, Beaverton, OR) and a custom MFP setup as described previously.[29] We used a system of neMESYS syringe pumps programmed using Qmix-Elements (Cetoni GmbH, Korbussen, Germany) to control the injection and aspiration flow rates in each channel. We synchronized the image acquisition with the sample illumination using a macro programmed in NIS-Elements software (Nikon Instruments, Melville, NY).

We used a 10× objective (CFI Plan Fluor Ph1 DLL, N.A. 0.3, Nikon Instruments, Melville, NY) with a FITC filter-cube (F36-525, AHF Analysetechnik, Tübingen, Germany) for flow-rate characterization and a 40× objective (CFI S-Plan Fluor ELWD DIC, N.A. 0.6, Nikon Instruments, Melville, NY) with a Cy3 HC filter-cube (F36-542, AHF Analysetechnik, Tübingen, Germany) for kinetic measurements.

All hybridization experiments were performed in an environmental chamber installed on the microscope (The Cube and The Box, Life Imaging Services, Basel, Switzerland), and set to a temperature of 37 °C to maintain a constant hybridization temperature.

## Reagents and Probes

To study the effects of different hybridization agents on the *in situ* hybridization kinetics, we set the baseline chemistry to consist of a probe concentration of 28 nM in a solution containing 150 mM NaCl (Sigma-Aldrich, St. Louis, MO), and varied in turn the probe concentration between 14 nM and 140 nM, and the ionic strength by adding between 150 mM and 1 M NaCl. To investigate the effects of volume exclusion and destabilizing agents on the hybridization, we used dextran sulfate (Molecular weight >500,000, Sigma-Aldrich, St. Louis, MO) at concentrations ranging between 0% and 10%, and formamide (Merck, Kenilworth, NJ) at concentrations ranging between 0% and 25%, respectively.

A list of the probe sequences used in the experiments, the thermodynamic considerations for the selection and determination of probe specificity, and a description of the MFP devices are all provided in the Supporting Information (SI), Section S-1-S-3.

## Cell Blocks

We used MCF-7 cell blocks purchased from AMS Biotechnology (T2255830, AMS, Abingdon, UK). We first baked the cell blocks for 1 hour at 55-60 ºC on a hotplate (VWR International, Randor, PA). Then we dewaxed the cell blocks in xylene (Merck, Kenilworth, NJ) in a glass beaker for 10 min, rinsed with 100% ethanol (Merck, Kenilworth, NJ) for 3 min and air-dried them at room temperature. We then immersed the dried blocks in an antigen-retrieval solution of 0.08×SSC (Thermo Fisher Scientific, Waltham, MA) for 20 min at 98 ºC, washed in DI for 5 min, and dried them at room temperature. We drew a hydrophobic barrier around the cells using a hydrophobic pen (Sigma-Aldrich, St. Louis, MO), and then digested the blocks with pepsin (Milan Analytica AG, Rheinfelden, Switzerland) for 8 min at 37 ºC, washed in DI for 5 min, and dried the cells at room temperature. We pipetted a FISH buffer (KI/KBI-FHB, Biosystems Switzerland AG, Muttenz, Switzerland) onto the cells, placed a coverslip on top, and let the cells denature for 5 min at 75 ºC. After denaturation, we immersed the cells in DI and gently removed the coverslip and washed the cells. We then transferred the slides to the microscope sample holder and pipetted 1×SSC (pre-warmed to 37 °C) onto the cells before the kinetic measurements.

**Kinetic Measurements**

We performed all kinetic measurements by switching between an injection of the probe solution and an imaging buffer composed of 1×SSC in DI. For each condition, we performed the measurements by repeating a cycle of (1) a 30-45 s injection of probes onto the surface, during which the hybridization reaction takes place, followed by (2) injection of a non-fluorescent imaging buffer, during which we captured the signal on the surface. We repeated the cycle until the fluorescent signal stabilized, indicating that the hybridization reaction reached a steady state. Once the signal had stabilized, we performed a final wash with 1×SSC for 5 min and verified that the fluorescent signal obtained is consistent with the steady-state measurements prior to the wash. The final wash served as a control, parallel to a detergent wash in an on-bench FISH assay.

In all measurements, the signal at each time point was captured by acquiring a z-stack of 5 images from the center to ±1 μm in height to account for the variations in the vertical position of the hybridization site within the nucleus.

**Post Processing and Image Analysis**

We analyzed the images in Matlab 2016b (MathWorks, Natick, MA), using a custom code (available upon request) that incorporates the selection of 3-5 FISH signals in the imaged cells based on their location in the final image (a detailed description of the analysis steps is provided in Section S-4 in the SI). Signal development over time was analyzed by averaging the FISH signal at the selected location of 12×12 pixels, corresponding to 2×2 μm over the area and z-stack at each time point. Each signal value was background-subtracted using the value of the background at steady state, and normalized by its steady state value calculated using fitting of the data to Eq. (3). The error bars were calculated based on a double-sided p-value of 0.05 (corresponding to 95% confidence on the mean), using at least 5 repeats..

# Theory and Principle of the Method

Our method is based on rapid switching between localized delivery of a fluorescent probe solution and of a non-fluorescent imaging buffer to the cells on a surface, which allows kinetic measurements of intracellular hybridization to be performed in real time.

Figure 1a-b presents the MFP head, which contains six microchannels: (1) 'probe injection', (2) 'probe removal', (3) 'buffer injection', (4) 'buffer removal', (5) 'injection', and (6) 'aspiration'. Channels (1)-(4) are connected to a 1×1 mm switching chamber located at a distance of 1.5 mm from the apex. The probe and buffer injection channels inject a probe solution or a non-fluorescent imaging buffer into the switching chamber, respectively. The probe and buffer removal channels are connected on each side on the bottom of the chamber and are designed to alternately aspirate the respective solution from the sides of the switching chamber. This configuration allows only the required solution to enter the injection channel, which is connected to the bottom of the chamber and delivers the solution to the surface. The solution is then aspirated through the aspiration channel, located at a distance of 100 μm from the injection, along with some of the surrounding immersion liquid. A higher aspiration flow rate than that of the injection allows a hydrodynamic confinement of the solution on the surface.

The removal flow rate in both the probe removal and buffer removal channels was set to a constant value of –7 nL min$^{-1}$. To initiate an injection of one of the solutions, its flow rate was set to 1 µL min$^{-1}$, whereas the flow rate of the other was set to the minimal value (1 nL min$^{-1}$). The constant minimal aspiration through the removal channels prevents the latter from reaching the injection channel, so that only the first solution is injected to the surface. The close proximity of the chamber to the surface and the continuous flow inside it are key for achieving rapid switching. A video showing the switching between the two solutions is provided in the SI (SM-1).

Figure 1d presents a schematic illustration of the switching scheme. The probe solution and a non-fluorescent imaging buffer are alternately delivered to the surface. Hybridization takes place only while probes are being delivered to the surface ($t_1$, $t_3$). During the imaging buffer injection phase ($t_2$, $t_4$), the intracellular FISH signal can be imaged directly on the surface because the background signal of unbound probes is eliminated. The total hybridization time is only the time over which probes are delivered to the surface. A video showing an example of FISH signal development over time is provided in the SI (SM-2).

## Switching Times

We characterized the transition time between two consequent solutions as the transition time between 10% and 90% of the maximum concentration on the surface. This transition is governed by Taylor–Aris dispersion and is given by [30]

$$t_s = \frac{4}{U\sqrt{U/LD_{eff}}} erf^{-1}(0.8)$$
$$D_{eff} = D\left(1 + \frac{\beta}{210}\frac{U^2 a^2}{D^2}\right) \quad , \tag{1}$$

where $U$ is the fluid velocity, $L$ is the length of the channel and $D_{eff}$ is the effective diffusion coefficient of the species, wherein $D$ is the molecular diffusion coefficient, $\beta$ is a geometrical dispersion coefficient, equal to 1.76 for a square cross section,[31] and $a$ is a cross sectional characteristic length, corresponding to the channel width.

A plot of the theoretical and experimental transition times as function of the flow rate in our geometry ($L = 2.5\,mm$, $a = 100\,\mu m$), assuming a diffusion coefficient of $2\times 10^{-11}\ m^2/s$,[32] is presented in Section S-5 in the SI.

## Considerations in Setting Flow Rates

When choosing the operating flow-rate we consider the trade-off between the transition time and stability of the signal on the surface. While high flow rates result in faster transition times, they also induce delamination of cells from the surface. In contrast, lower flow rates result in a noisy signal on the surface owing to the higher sensitivity of the system to disturbances. Thus, the operating flow rate chosen (1 µL/min) is the result of a trade-off between the two which allows a relatively fast switching between reagents ($t_s = 10.57\ s$), while cell adherence remains intact.

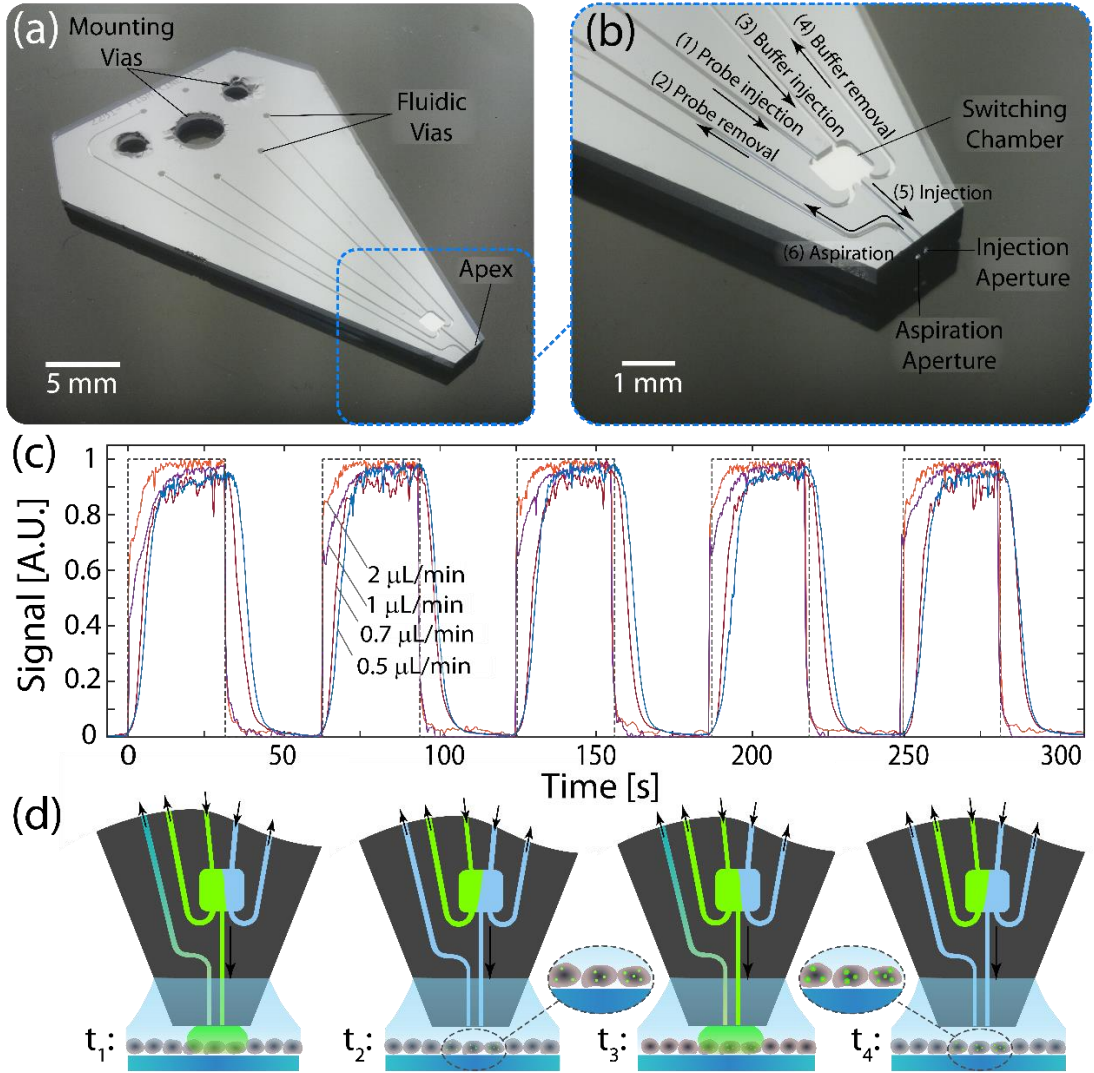

*Figure 1.* MFP device, principle of the method, and experimental conditions. (a-b) Photograph of the device and channel geometry. Two pairs of 100-μm-wide and 100-μm-deep microchannels are connected to a 1×1 mm switching chamber. The inner channel in each pair is designed to inject a non-fluorescent imaging buffer solution or probe solution, respectively, whereas the outer channel is designed for removal of the respective solution. The flow rates of the imaging buffer and probe solution were alternately switched between the flow rate chosen and 0, whereas the removal flow rate was set to 7 nL/min. (c) Characterization of the fluorescent signal on the surface for flow rates of 2, 1, 0.7 and 0.5 μL/min. For flow rates of 1–2 μL/min, the signal is stable and follows the input function closely (transition time from 10% to 90% of the maximal signal of 5.43 s and 10.57 s, respectively). Flow rates below 1 μL/min lead to a more substantial switching time between the probe solution and the imaging buffer (transition time of 13.4 s and 15.75 s for 0.7 μL/min and 0.5 μL/min, respectively). (d) Schematic illustration of the switching scheme and real-time imaging of the FISH signal development. The probe solution and a non-fluorescent imaging buffer are alternately delivered to the surface. While probes are being delivered to the surface ($t_1$; $t_3$) hybridization takes place; during the imaging buffer phase ($t_2$; $t_4$) the intracellular FISH signal can be imaged directly on the surface because the background signal of unbound probes is removed.

**Hybridization Kinetics**

The hybridization reaction of the free probes to chromosomal targets in the nucleus can be described using second-order reaction kinetics,[33]

$$\frac{\partial c_{PT}}{\partial t} = k_{on} c_P c_T - k_{off} c_{PT} ,  \qquad (2)$$

where $c_P$ is the concentration of the probe, $c_T$ is the concentration of target in the nucleus, $c_{PT}$ is the concentration of bound probes, $k_{on}$ and $k_{off}$ are, respectively, the on- and the off-rate of the hybridization reaction. Assuming the probe concentration is much higher than the target, $c_P \gg c_T$, the concentration of the bound probes as function of time can be described as

$$c_{PT} = c_T \left(1 + K_d / c_P\right)\left(1 - \exp\left[-\left(k_{on} c_P + k_{off}\right) t\right]\right) ,  \qquad (3)$$

where $K_d$ is the dissociation constant, $K_d = k_{off} / k_{on}$, and $t$ is the reaction time. The dissociation constant can also be expressed as $K_d = \exp(-\Delta G / RT)$, where $\Delta G$ is the Gibbs free energy of the hybridization reaction, $R$ is the universal gas constant and $T$ is the temperature.

The thermodynamic considerations for the selection and determining of probe sequences and their specificity are provided in detail in Section S-4 in the SI.

# Results and Discussion

The hybridization efficiency and specificity in a FISH assay depend on multiple variables including the composition of the hybridization buffer, the probe concentration, the hybridization temperature, and the probe sequence itself. Here we demonstrate the applicability of our method for the characterization of the FISH reaction under conditions of varying probe concentration, volume exclusion agent (dextran sulfate) content and ionic strength conditions, using a set of centromeric probes (Cen17). An additional example showing the effects of a destabilizing agent (formamide) content is provided in Section S-6 in the SI. We show that our method can be used to investigate the effect of each of these variables and provide insight into processes affecting *in situ* hybridization.

**Probe Concentration Dependence**

Figure 2a presents kinetic measurements of a FISH reaction at increasing probe concentrations from 14 nM to 140 nM as function of $t \cdot C_0$. As follows from Equation 3, one would expect that for an ideal first-order reaction, all experiments will collapse to a single curve indicating the on-rate. However, we see that whereas the probe concentrations of 28 and 70 nM yield the same on-rate, a higher on-rate is obtained at a lower concentration (14 nM) and a lower one is obtained at a higher concentration (140 nM). At low concentrations, there is likely minimal electrostatic interaction between the probe sequences, enabling a quicker interaction with the targets, explaining the slightly higher on-rate for 14 nM than for 28 and 70 nM. For a probe concentration of 140 nM, we observe a ~1.5-fold reduction of the hybridization rate compared with the other concentrations, to $6.55 \times 10^4$ M$^{-1}$s$^{-1}$. We hypothesize that this decrease is due to either inter-binding of the probes (if $c_P \gg K_d$ probe-to-probe) or to crowding effects near the target sites.

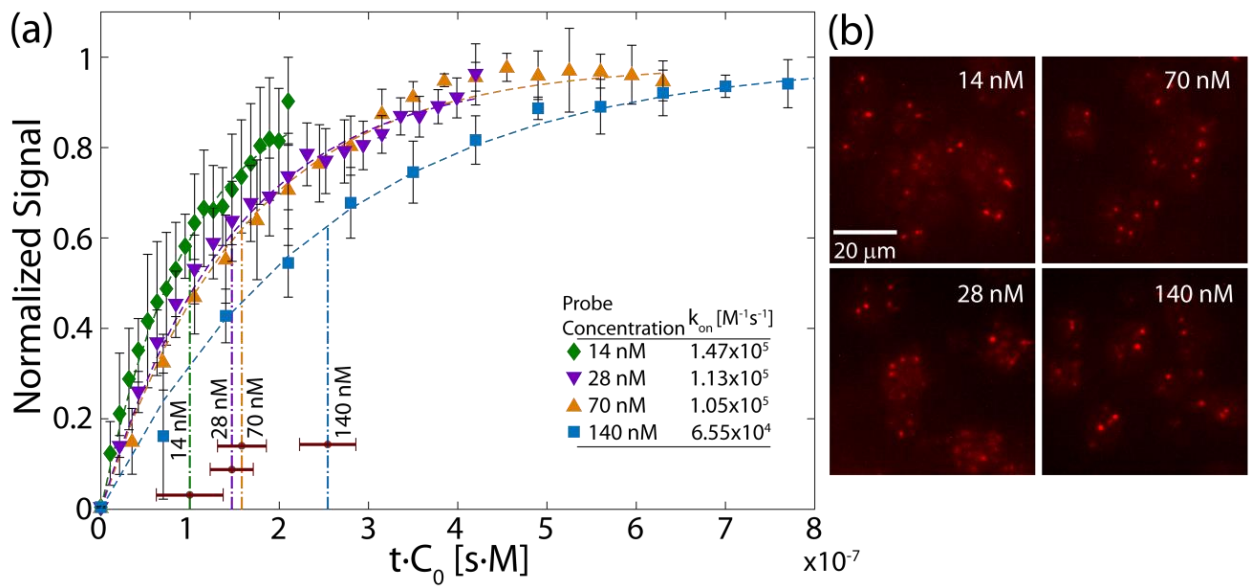

*Figure 2.* Experimental results showing kinetic measurements for different probe concentrations as a function of $t \cdot C_0$. (a) Hybridization curves for probe concentrations of 14 nM, 28 nM, 70 nM, and 140 nM. For concentrations between 14 and 70 nM, the measured on-rate remains practically unchanged and the curves collapse to a single curve as function of $t \cdot C_0$, as expected. For a probe concentration of 140 nM, the on-rate decreases because of a transition to a reaction-limited hybridization mode. Vertical lines and red horizontal error bars correspond to $k_{on}^{-1}$ of each reaction. All error bars correspond to 95% confidence on the mean using at least 5 repeats. (b) Fluorescence images of end-point results of the FISH hybridization reaction for a group of cells at each of the concentrations after 15 min of incubation. FISH signals exhibit similar intensities at all concentrations after the reaction has reached saturation. All images are background subtracted.

The average value of $k_{on}$ for probe concentrations between 14 and 70 nM was $1.22 \times 10^5$ M$^{-1}$s$^{-1}$, having a range of $1.05 \times 10^5$ M$^{-1}$s$^{-1}$ and $1.47 \times 10^5$ M$^{-1}$s$^{-1}$. This rate is more than twice lower than surface-based kinetic measurements of standard 20-nt-long DNA targets hybridizing to immobilized probes of the same length, and having a similar GC content of 50%, using graphene [34] and SPR-based [35] sensors, which resulted in on-rates of $2.58 \times 10^5$ $M^{-1}s^{-1}$ and $2.90 \times 10^5$ $M^{-1}s^{-1}$ respectively. We attribute this decrease in $k_{on}$ to intracellular and nuclear molecular crowding effects leading to a slower transport of probes as well as to limited accessibility of the target sites due to the spatial arrangement of the genomic DNA. [16,18,19] Our measured on-rates are in agreement with measurements by Rode *et al.*[36] of hybridization of DNA probes of similar length (15 nt) to cell-surface immobilized targets, which resulted in rates ranging between $1.03 \times 10^5$ and $1.93 \times 10^5$ $M^{-1}s^{-1}$. Similarly, the lower rates in that study were explained by limited target access due to crowding effects, steric hindrance, and electrostatic repulsion from the anionic glycocalyx.[36] Whereas in FISH cells are partially digested and thus the extent of molecular crowding is reduced relative to intact cells, hybridization of probes to genomic DNA takes place inside the nucleus, and thus target accessibility due to DNA conformation and additional nuclear crowding have to be taken into account.[16,18,19]

**Volume Exclusion Agents (Dextran Sulfate)**

Volume exclusion agents are polymers commonly used in *in situ* hybridization assays to enhance the hybridization rates by creating a crowded environment in which the polymer occupies solvent space, and increases the local concentration of DNA.[33,37–39] We demonstrate the applicability of our method to investigate effects of volume exclusion agents by characterizing the hybridization rates in hybridization buffers containing varying concentrations of dextran sulfate. Dextran sulfate is considered the most suitable volume exclusion agent for *in situ* hybridization because of its ability to significantly accelerate hybridization kinetics, while at the same time retaining low levels of non-specific signals.[40] Figure 3 presents kinetic FISH measurements for hybridization buffers containing 0%–10% dextran sulfate. The hybridization rate increases linearly with increasing dextran sulfate content as shown in Figure 3b. Higher concentrations of dextran sulfate resulted in too high a viscosity of the hybridization buffer which could not be confined using the MFP.

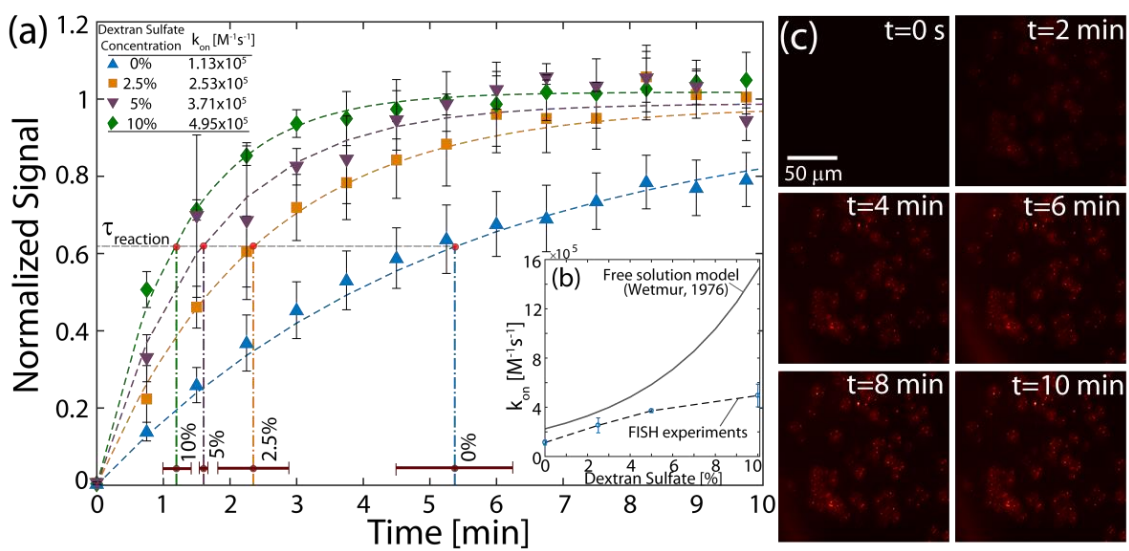

*Figure 3. Experimental results showing kinetic measurements for different concentrations of dextran sulfate as function of time. (a) Hybridization curves for 28 nM probe concentrations in hybridization buffer containing 0%, 2.5%, 5% and 10% (v/v) dextran sulfate. (b) A plot of $k_{on}$ vs. dextran sulfate content obtained in our experiments Dashed lines show trend lines to guide the eye. Continuous grey curve shows the expected rate increase according to the free solution model.[37] (c) A time sequence of a group of cells undergoing in situ hybridization in a buffer containing 5% dextran at different points in time. The signal increases over time until saturation. The scale bar refers to all images; all images are background subtracted.*

There is anincrease in FISH hybridization rate with increasing dextran sulfate content. The increase is less pronounced than in free solution, in which an exponential relation between the content of dextran sulfate and the renaturation rate of DNA was reported.[33] However, the observed change in the hybridization rate in FISH is far more pronounced than in live cells, in which molecular crowding effects are stronger and thus exclusion agents were shown to have a minimal effect on the intracellular hybridization rate.[15] This trend implies that while the effects of molecular crowding are reduced in FISH owing to digestion of the cellular and nuclear matrix during the cell pre-treatments, these effects are not entirely lost, leading to a more moderate effect on the hybridization rates than in the bulk. Furthermore, in FISH, the volume exclusion agent mainly affects the free probes, while the targets are immobile in the cell. Thus the effective concentration of the probe, $c_p$, increases, while the target concentration, $c_t$, remains unchanged.

**Ionic Strength**

The ionic strength strongly affects DNA hybridization by stabilizing the helix structure of both single-stranded DNA and duplexes,[41] and neutralizing electrostatic repulsion between complementary strands through shielding of the negatively charged phosphate backbone.[42,43]

In FISH reactions, salt is used to stabilize the hybridized nucleic acid duplexes by serving two functions: stabilizing the DNA strands, similarly to hybridization in solution,[44,45] and compensating for the increased negative charge density in the nucleus.[46,47] The ionic strength dependence in a FISH reaction is thus expected to reflect both of these effects. The latter is parallel to surface hybridization assays, in which the immobilized probes create a negative charge density layer.[48,49] This layer leads to an electrostatic barrier to hybridization, implying that at low ionic strength conditions, hybridization would not occur unless solution ions provide sufficient charge screening. Equivalently, owing to the high concentration of negative charges of the densely packed DNA in the nucleus, we expect a similar electrostatic barrier to be present. Figure 4a presents kinetic measurements of FISH hybridization at salt concentrations varying between no salt and 1 M NaCl. With no salt present, as expected, no hybridization occurred, indicating that the binding free energy is too low. Figure 4b presents the measured $k_{on}$ as function of the ionic strength. The hybridization rate increases with increasing salt concentration. We identify two regimes in the plot: whereas a prominent rate increase occurs as the concentration is increased to 300 mM NaCl, the rate increase is more moderate as we further increase the salt concentration up to 1 M. We attribute the steeper slope at lower ionic strength to a combined effect of duplex stabilization, and electrostatic shielding of the nuclear negative charge. The effectivity of the latter rapidly increases at lower ionic strengths (<300 mM), owing to a higher activity coefficient, according to the Debye–Hückel model.[50] As the salt concentration is further increased, the Debye-Hückel model is no longer valid, a transition to a different behavior occurs owing to steric effects and the rate increase becomes more moderate. These results are consistent with measurements by Owczarzy et al.,[51] and Lang et al.[52]

Figure 4c presents fluorescence images of end-point results at different ionic strengths. At high ionic strength of 1 M NaCl, we observe a relatively high level of off-target signal in the nucleus. These non-specific signals surrounding the (brighter) specific signals lead to variability in the mean signal in the analyzed region of interest (ROI), as described in the Experimental Section, causing the larger error bars obtained at this condition.

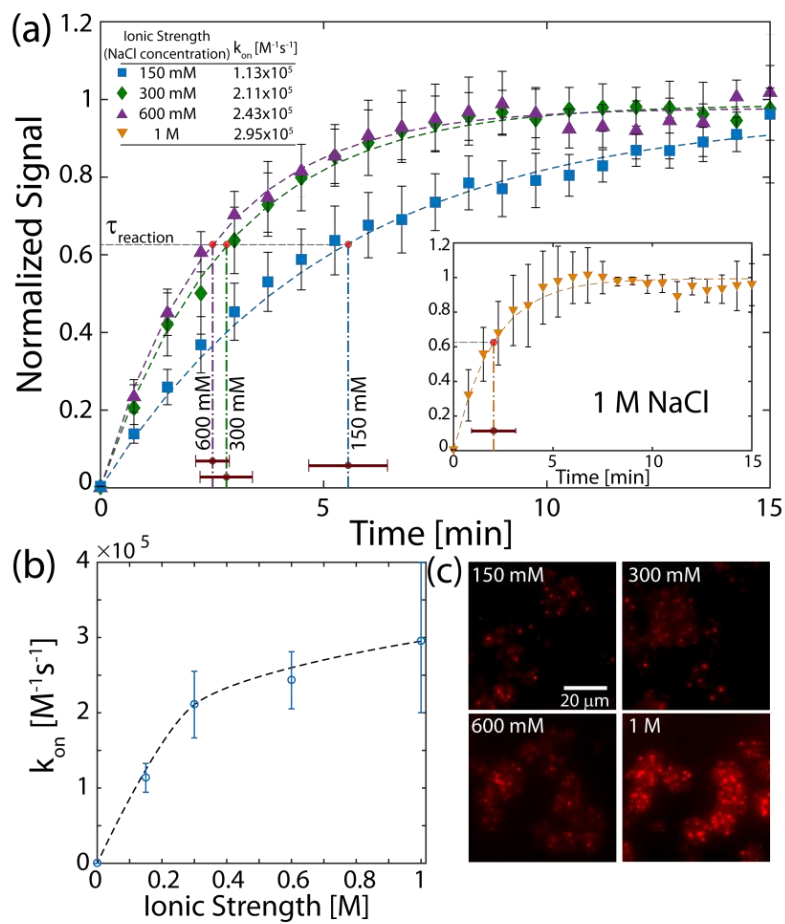

*Figure 4. Experimental results showing kinetic measurements at varying ionic strength of the hybridization buffer as function of time. (a) Hybridization curves for 28 nM probe concentrations in a hybridization buffer containing 150 mM, 300 mM, 600 mM and 1 M NaCl. (b) Plot of $k_{on}$ as function of the ionic strength. Whereas a notable increase in $k_{on}$ occurs as the ionic strength is increased to 300 mM, the increase is more moderate as the ionic strength is further increased. The dashed line shows a trend line to guide the eye. (c) Fluorescence images of end-point results at each of the ionic strength conditions after 15 min of incubation. The scale bar refers to all images; all images are background subtracted.*

## Conclusions

We present a new method for the real-time monitoring and kinetic analysis of the FISH signals. We combine convection-enhanced transport and rapid switching between a fluorescent probe solution and a non-fluorescent imaging buffer. The FISH signal is recorded only during the imaging-buffer phase, thus, while imaging, the background signal associated with unhybridized probes is completely removed and accurate kinetic measurements can be obtained in real time. We implemented the method using a microfluidic probe, which is beneficial owing to its non-contact operation on surfaces, allowing interaction with confined on-surface regions without physical disruption of the sample. The convection-based delivery of constant probe and solution concentrations to the surface allows kinetic measurements to be performed in relatively short times, in contrast to the typical diffusion-based delivery in FISH assays, associated with long assay times, and which allow only a single end-point measurement per assay, after a stringent wash.

We demonstrated the applicability of the method for monitoring the effects of four hybridization conditions typically used to increase the *in situ* hybridization efficiency: probe concentration, volume exclusion agents content (dextran sulfate), destabilizing agents content (formamide), and ionic strength. We note that our method is not limited to the effects demonstrated here and can be also used to characterize effects of other conditions and solvents, for example, temperature, probe sequences, lengths and types (e.g., peptide nucleic acid or morpholino probes), and other rate-enhancing agents (e.g., ethylene carbonate, sulfolane, or γ-butyrolactone).[24] We demonstrated kinetic measurements of the effective on-rate of a probe mix, as FISH probes are typically a mix of sequences. The effective rates of different mixes as well as the on-rate of a single sequence can equivalently be monitored and compared using our method.

While in this work we performed separate kinetic measurements to analyze the effects of each condition and the off-rate in our reaction was negligible, we note that the method can be also used to monitor combinatory effects as well as dissociation reactions when $k_{off}$ is significant, by constantly delivering only the imaging buffer after the hybridization reaction has reached saturation and monitoring the signal as the probes dissociate. Furthermore, the method can be used to monitor reactions and binding events on the cell membrane.

The experimental results presented in this work show that the *in situ* hybridization reaction cannot be accurately described by models based on hybridization in cell-free systems and that unique effects for FISH, such as molecular crowding and the local charge density, should be taken into account to construct a comprehensive theoretical or computational model. Such effects could not be observed previously because FISH assays generally rely on diffusion-based transport of probes, leading to uncontrolled concentrations on the surface.

We believe this method opens the door to new insights and a better understanding of *in situ* reaction processes which could not be formerly quantified owing to the limitations imposed by the traditional FISH techniques, and will facilitate the design of new assays and protocols as well as more accurate theoretical or computational models describing *in situ* hybridization.

## Author Information

**Corresponding Author**
* E-mail: gov@ibm.zurich.com
**Author Contribution**
§These authors contributed equally.
**Notes**
The authors declare no competing financial interest.

## Acknowledgements

This work was supported by the ITN-EID grant, under the 7[th] Framework Program (Project No. 607322, Virtual Vials), and by the European Research Council (ERC) Starting Grant, under the 7th Framework

Program (Project No. 311122, BioProbe). We thank Baruch Rofman for valuable discussions. Dr. Emmanuel Delamarche and Dr. Walter Riess are acknowledged for their continuous support.
# References

(1) Cassidy, A.; Jones, J. Developments in in Situ Hybridisation. *Methods* **2014**, *70* (1), 39–45.
(2) Wagner, M.; Horn, M.; Daims, H. Fluorescence in Situ Hybridisation for the Identification and Characterisation of Prokaryotes. *Curr. Opin. Microbiol.* **2003**, *6* (3), 302–309.
(3) Summersgill, B.; Clark, J.; Shipley, J. Fluorescence and Chromogenic in Situ Hybridization to Detect Genetic Aberrations in Formalin-Fixed Paraffin Embedded Material, Including Tissue Microarrays. *Nat. Protoc.* **2008**, *3* (2), 220–234.
(4) Kallioniemi, A.; Kallioniemi, O. P.; Sudar, D.; Rutovitz, D.; Gray, J. W.; Waldman, F.; Pinkel, D. Comparative Genomic Hybridization for Molecular Cytogenetic Analysis of Solid Tumors. *Science* **1992**, *258* (5083), 818–821.
(5) Tapia, C.; Glatz, K.; Novotny, H.; Lugli, A.; Horcic, M.; Seemayer, C. A.; Tornillo, L.; Terracciano, L.; Spichtin, H.; Mirlacher, M.; et al. Close Association between HER-2 Amplification and Overexpression in Human Tumors of Non-Breast Origin. *Mod. Pathol.* **2007**, *20* (2), 192–198.
(6) Copie-Bergman, C.; Gaulard, P.; Leroy, K.; Briere, J.; Baia, M.; Jais, J.-P.; Salles, G. A.; Berger, F.; Haioun, C.; Tilly, H. Immuno–Fluorescence In Situ Hybridization Index Predicts Survival in Patients With Diffuse Large B-Cell Lymphoma Treated With R-CHOP: A GELA Study. *J. Clin. Oncol.* **2009**, *27* (33), 5573–5579.
(7) Amann, R.; Fuchs, B. M. Single-Cell Identification in Microbial Communities by Improved Fluorescence in Situ Hybridization Techniques. *Nat. Rev. Microbiol.* **2008**, *6* (5), 339–348.
(8) Sarrate, Z.; Anton, E. Fluorescence in Situ Hybridization (FISH) Protocol in Human Sperm. *J. Vis. Exp.* **2009**, No. 31.
(9) Hattab, E. M.; Martin, S. E.; Al-Khatib, S. M.; Kupsky, W. J.; Vance, G. H.; Stohler, R. A.; Czader, M.; Al-Abbadi, M. A. Most Primary Central Nervous System Diffuse Large B-Cell Lymphomas Occurring in Immunocompetent Individuals Belong to the Nongerminal Center Subtype: A Retrospective Analysis of 31 Cases. *Mod. Pathol.* **2010**, *23* (2), 235.
(10) Pattnaik, P. Surface Plasmon Resonance. *Appl. Biochem. Biotechnol.* **2005**, *126* (2), 79–92.
(11) Ghai, R.; Falconer, R. J.; Collins, B. M. Applications of Isothermal Titration Calorimetry in Pure and Applied Research-Survey of the Literature from 2010: SURVEY OF RECENT APPLICATIONS OF ITC. *J. Mol. Recognit.* **2012**, *25* (1), 32–52.
(12) Khanna, V. K. *Nanosensors: Physical, Chemical, and Biological*; CRC Press, 2011.
(13) Hancock, R.; Jeon, K. W. *New Models of the Cell Nucleus: Crowding, Entropic Forces, Phase Separation, and Fractals*; Academic Press, 2013; Vol. 307.
(14) Wienberg, J.; Stanyon, R. Comparative Painting of Mammalian Chromosomes. *Curr. Opin. Genet. Dev.* **1997**, *7* (6), 784–791.
(15) Schoen, I.; Krammer, H.; Braun, D. Hybridization Kinetics Is Different inside Cells. *Proc. Natl. Acad. Sci. U. S. A.* **2009**, *106* (51), 21649–21654.
(16) Annunziato, A. DNA Packaging: Nucleosomes and Chromatin. *Nat. Educ.* **2008**, *1* (1), 26.
(17) Buenrostro, J. D.; Wu, B.; Litzenburger, U. M.; Ruff, D.; Gonzales, M. L.; Snyder, M. P.; Chang, H. Y.; Greenleaf, W. J. Single-Cell Chromatin Accessibility Reveals Principles of Regulatory Variation. *Nature* **2015**, *523* (7561), 486–490.
(18) Wedemeier, A.; Merlitz, H.; Wu, C.-X.; Langowski, J. Modeling Diffusional Transport in the Interphase Cell Nucleus. *J. Chem. Phys.* **2007**, *127* (4), 45102.
(19) Spencer, V. A.; Xu, R.; Bissell, M. J. Extracellular Matrix, Nuclear and Chromatin Structure, and Gene Expression in Normal Tissues and Malignant Tumors: A Work in Progress. In *Advances in Cancer Research*; Elsevier, 2007; Vol. 97, pp 275–294.
13